\documentclass[showpacs, oneside, twocolumn, prl, amsmath, amssymb, nofootinbib, superscriptaddress]{revtex4-1}

\usepackage{cases}
\usepackage{amsmath}
\usepackage{amssymb}
\usepackage{amsfonts}
\usepackage{amssymb}
\usepackage{dcolumn}
\usepackage{bm}
\usepackage{bbm}
\usepackage{graphicx}
\usepackage{xcolor}
\usepackage{array}
\usepackage{subfigure}
\usepackage{hyperref}
\usepackage{wasysym}

\newcommand{\be}{\begin{equation}}
\newcommand{\ee}{\end{equation}}
\newcommand{\ba}{\begin{eqnarray}}
\newcommand{\ea}{\end{eqnarray}}

\newcommand{\gsim}{\mathrel{\hbox{\rlap{\lower.55ex \hbox {$\sim$}}
                   \kern-.3em \raise.4ex \hbox{$>$}}}}
\newcommand{\lsim}{\mathrel{\hbox{\rlap{\lower.55ex \hbox {$\sim$}}
                   \kern-.3em \raise.4ex \hbox{$<$}}}}

\hypersetup{colorlinks=true,
            breaklinks=true,
            pdfstartview=Fit,
            linkcolor=blue,
            citecolor=blue,
            urlcolor=blue}

\begin{document}
\title{A Bit Rate Bound on Superluminal Communication}

\author{Xi Tong}
\email{xtongac@connect.ust.hk}

\author{Yi Wang}
\email{phyw@ust.hk}

\author{Yuhang Zhu}
\email{yzhucc@connect.ust.hk}

\affiliation{Department of Physics, The Hong Kong University of Science and Technology, Clear Water Bay, Kowloon, Hong Kong, P.R.China}
\affiliation{The HKUST Jockey Club Institute for Advanced Study, The Hong Kong University of Science and Technology, Clear Water Bay, Kowloon, Hong Kong, P.R.China}

\begin{abstract}
We study semi-classical communication in positivity-violating k-essence scalar field theories, with superluminal modes propagating on a rolling background. The self-interactions due to the non-linear nature of these theories pose a constraint on the rate of superluminal information transfer. We derive a novel bit rate bound on superluminal communication within a conceptual model, to which a general class of k-essence theories naturally reduces. Our result implies the possibility that, even if these positivity-violating k-essence theories may not possess a maximal information propagation speed, there is nevertheless an upper bound on the rate of information transfer.
\end{abstract}

%\pacs{98.80.Cq, 11.25.Tq, 74.20.-z, 04.50.Gh}

\maketitle

%\section{Introduction}
{\it Introduction.} --
Special relativity \cite{Einstein:1905ve,Einstein:1907iag} is undoubtedly one of the most precise and concise theories of physics. Arising from its two underlying principles, is a profound spacetime symmetry known as Lorentz invariance (LI). The causal structure of spacetime compatible with LI thus prohibits any superluminal motion that carries information \cite{Brillouin:1960tos,Fox:1969us,1966Natur.211..468R,Davis:2003ad}, and the speed of light in the vacuum $c=1$ becomes the universal upper bound for information propagation. When consistently quantized, LI also puts tight constraints on the matter sector \cite{1939AcHPh..12....3F,Pauli:1940zz,Luders:1954zz,Weinberg:1964ew,Coleman:1967ad,Weinberg:1980kq}. Experiment-wise, decades of search for signs of Lorentz violation has only succeeded in improving its accuracy \cite{Colladay:1998fq,Coleman:1998ti,Kostelecky:2008ts,Cohen:2011hx,Liberati:2013xla}. Despite the great success of special relativity, it remains a possibility, if not an elegant one by today's standards, that there exists a Lorentz-violating dark sector so weakly coupled to the Standard Model (SM) that all the present experiments have not yet unveiled. Without the constraint of LI, the speed limit may be alleviated, possibly even indefinitely.

It has been known for a long time that certain families of scalar field theories admit superluminal modes when expanded upon an IR background \cite{ArmendarizPicon:1999rj,Garriga:1999vw,ArmendarizPicon:2000dh,ArmendarizPicon:2000ah,Mukhanov:2005bu}, which spontaneously breaks LI (for similar superluminality setups in different contexts, see \cite{Drummond:1979pp,Scharnhorst:1990sr,Barton:1989dq,Jacobson:2004ts}). These k-essence scalars, although equipped with a formally local and Lorentz-invariant Lagrangian, are secretly non-local macroscopically \cite{Adams:2006sv}. The wrong signs in front of the higher-derivative operators in their Lagrangian forbid any UV completion satisfying the usual axioms of S-matrix theory. However, an open mind towards this is that, gravity aside \cite{Shore:2007um}, these unusual IR superluminal k-essence theories may not need to have a usual UV completion satisfying locality, analyticity, unitarity and LI. They stand on their own as consistent IR theories respecting a weaker notion of causality \cite{ArmendarizPicon:2005nz,Bruneton:2006gf,Babichev:2007dw}, with a novel UV description unlike anything we have developed before. Therefore, it is nevertheless useful to understand their novelty within the IR region as a first step, namely, the exciting possibility of superluminal communication.

The reason for such a study is two-fold. On one hand, superluminal modes, if exist, can be used to transmit information, which can be technologically useful. For instance, the k-essence field may live in the dark sector as a potential dark ``Aether'' for communication. The protocol for this superluminal-communication channel is thus necessary. On the other hand, if indeed a universal \textit{speed} bound for information transfer is relaxed and locality/LI is forsaken, nature might still pose a universal \textit{bit rate} bound for information transfer.
%, as a restriction to the complexity of the evolution of universe.
As we will show in this \textit{Letter}, a bound on the bit rate of sending information can be derived from inspecting the behavior of superluminal k-essence theories in the IR.

%\section{k-essence and superluminality}
{\it k-essence and superluminality.} --
We begin with a lighting review of superluminality in k-essence theories \cite{ArmendarizPicon:1999rj,Garriga:1999vw,ArmendarizPicon:2000dh,ArmendarizPicon:2000ah,Mukhanov:2005bu}. Consider a scalar field (the k-essence) protected by a shift symmetry $\phi\to\phi+\mathrm{const}$. The Lagrangian $\mathcal{L}(X)$ is purely a function of $X=-\frac{1}{2}(\partial\phi)^2$. The Equation of Motion (EoM) of $\phi$ is
\begin{equation}
G^{\mu\nu}\nabla_\mu\nabla_\nu \phi=0~,
\end{equation}
where $\nabla_\mu$ is the covariant derivative compatible with the original spacetime metric $g_{\mu\nu}$, while small fluctuations around a fixed background \cite{Wald:1984rg} live effectively on an emergent geometry \cite{Barcelo:2005fc} defined by
$
	G^{\mu\nu} = c_s \mathcal{L}_{,X}^{-2} (\mathcal{L}_{,X}g^{\mu\nu}-\mathcal{L}_{,XX}\nabla^\mu\phi\nabla^\nu\phi) ~,
$
with sound speed
\begin{equation}
	c_s^{-2}=1+2X\frac{\mathcal{L}_{,XX}}{\mathcal{L}_{,X}}~.\label{csDef}
\end{equation}
This is most easily seen by expanding $\phi=\phi_0+\varphi$, where $\phi_0$ is a solution to the EoM of on-shell evolution. The perturbation evolves as
\begin{equation}
G^{\mu\nu}D_\mu D_\nu\varphi=0~.\label{kessencepertEoM}
\end{equation}
Here $D$ is the covariant derivative calculated from the emergent metric $G$.
 
The sound speed $c_s$ can be either subluminal or superluminal, depending on the sign of $\mathcal{L}_{,XX}$. The subluminal branch may enjoy a conventional UV completion. The superluminal branch, however, is generally believed to have no traditional UV completions in the absence of gravity \cite{Shore:2007um}, because it violates the positivity bound, which is a direct consequence of locality, analyticity, unitarity and LI in the UV \cite{Pham:1985cr,Adams:2006sv,Nicolis:2009qm,deRham:2017avq}. Therefore, superluminality in the IR must stem from the highly non-trivial UV physics. Although $\varphi$ travels outside the light cone, a restricted notion of causality is still respected under mild conditions of the background. Namely, there is a global time function associated with the rest frame of the background and therefore the IR EFT can be made free of paradoxical closed causal curves \cite{Babichev:2007dw}.

A complete theory of k-essence also includes non-linear terms. One useful way to introduce non-linearity is by \textit{defining} the EFT as a theory only for perturbations in $X$. Namely, the EFT tower is built around a specific classical background:
\begin{equation}
\mathcal{L}(X)=\sum_{n=0}^{\infty}\frac{1}{n!}\partial_X^{n}\mathcal{L}(X_0)(X-X_0)^n~,\label{PertEFT}
\end{equation}
where $\partial_X^{n}\mathcal{L}\equiv\frac{c_n}{\Lambda^{4n-4}}$. The validity of such an expansion is controlled by the energy of the perturbations only.

As a further simplification, we will study the rigid limit of Minkowski spacetime and decouple gravity by sending $M_p\to\infty$. Taking $\phi_0=\dot\phi_0 t$ and $X_0=\frac{\dot\phi_0^2}{2}$, we then collect the leading interactions for $\varphi$:
\begin{equation}
	\mathcal{L}(X)=\text{(total derivatives)}+\mathcal{L}^{(2)}+\mathcal{L}_{\text{int}}~,
\end{equation}
with
\begin{eqnarray}
\mathcal{L}^{(2)}&\equiv&\frac{1}{2}\left[\left(c_1+\frac{c_2 \dot\phi_0^2}{2\Lambda^4}\right)\dot{\varphi}^2-c_1(\nabla\varphi)^2\right]\\
\mathcal{L}_{\text{int}}&\equiv&\left(\frac{c_2 \dot\phi_0}{2\Lambda^4}+\frac{c_3 \dot\phi_0^2}{6\Lambda^8}\right)\dot{\varphi}^3-\frac{c_2 \dot\phi_0}{2\Lambda^4}\dot{\varphi}(\nabla\varphi)^2+\cdots~.\label{expandedLagrangian}
\end{eqnarray}
Thus the sound speed is modified according to (\ref{csDef}). Under the large-$c_s$ limit, the leading non-linearity comes from the $\dot{\varphi}^3$ term, with a coupling constant naturally\footnote{Of course, the coupling constant quantitatively depends on another parameter $c_3$, but adjusting $c_3$ to cancel the contribution of $c_2$ would be fine-tuning. Thus based on the naturalness of the Wilson coefficients, we regard the $c_2$ term to be the dominant non-linearity source, up to an $\mathcal{O}(1)$ constant factor.} set by $\frac{\eta}{\dot\phi_0}$, $\eta\equiv\frac{c_2 \dot\phi_0^2}{2\Lambda^4}$.

%\section{A bit rate bound}
{\it A bit rate bound.} --
Such superluminality of perturbations is in conflict with the notion that $c=1$ is the maximal speed of information transfer. Furthermore, the information propagation speed $c_s$ increases indefinitely with stronger background field strength while maintaining causality. Naively, there appears to be no upper limit to our ability to transfer information, because neither the speed or capacity of the flow of information seems to receive any restriction.

This is the case if one only considers the quadratic part $\mathcal{L}^{(2)}$, since it gives rise to a linear wave equation without dispersion and solutions describing a trivial shape-preserving translation, $\varphi=\varphi(x\mp c_s t)$. However, one should be reminded that superluminality is a feature of the k-essence model (\ref{PertEFT}) as a whole. In the complete picture, the seemingly unlimited ability of communication is weakened. For even if we can consistently let $c_s\to\infty$, we still run into the inevitable self-interaction of the perturbations whose intensity grows with the information content. This fact turns out to set an upper bound on the bit rate of superluminal communication. As we will demonstrate below, given the ability to perform superluminal communication with unlimited speed, one still has a limited bit rate of information transfer.

We consider the large-$c_s$ limit. The leading contribution of superluminal $c_s$ is naturally associated with the interaction $\mathcal{L}_{\text{int}}\supset\frac{\eta}{\dot\phi_0}\dot{\varphi}^3$. We require it to be smaller than the quadratic Lagrangian:
\begin{equation}
\mathcal{L}_{\text{int}}\sim\left|\frac{\eta}{\dot\phi_0}\dot{\varphi}^3\right|<\mathcal{L}^{(2)}\sim\frac{1}{2}\left(1+\eta\right)\dot{\varphi}^2~,
\end{equation}
where we have taken $c_1\approx 1$. This translates into a constraint on the perturbation field gradient,
\begin{equation}
	|\dot{\varphi}|\lesssim\frac{\dot\phi_0}{2 c_s^2}~,\label{ImportantConstraint}
\end{equation} 
which, in turn, translates into a constraint on the energy-momentum tensor of free perturbations,
\begin{equation}
|T_{\mu\nu}^{(2)}|\sim c_s^{-1}\dot{\varphi}^2\lesssim \frac{\dot\phi_0^2}{c_s^5}~,
\end{equation}
where we have neglected all $\mathcal{O}(1)$ constants. In the classical theory, a constraint on the energy flux is all there is to say about the system. However, upon quantization, the constraint on the energy flux automatically becomes a constraint on the information flux, since the energy of the signal is quantized in units of massless field quanta, which carries the encoded information. 

For concreteness, let us set up a toy model for superluminal communication using k-essence perturbations. As a leading-order approximation, the non-linearity will be turned down except when we evoke the constraint (\ref{ImportantConstraint}). Before discussing the details, we first point out that our communication model is essentially semi-classical in the sense that it does not entail the entanglement of quantum states. The information carried by the perturbations $\varphi$ resembles a classical analog signal carried, for example, by electromagnetic waves transmitted between a pair of parabolic dish antenna in point-to-point telecommunications.

\begin{figure}[h!]
	\centering
	\includegraphics[width=8cm]{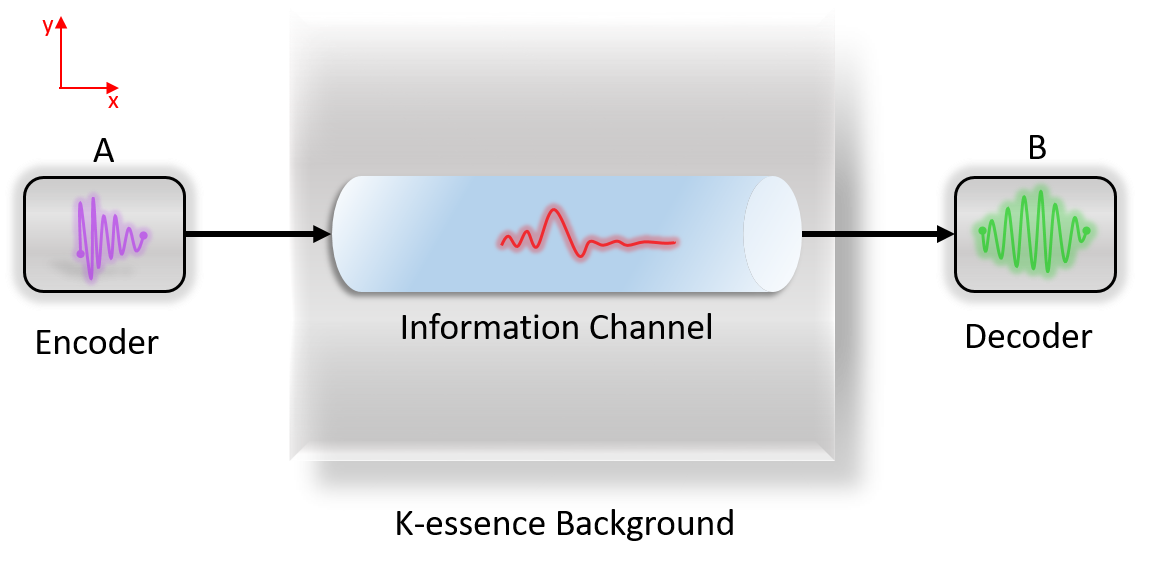}
	\caption{An illustration of our superluminal communication model.}\label{SignalingIllustration}
\end{figure}

Suppose there are two local observers (A and B) separated in the $x$-direction (see FIG.~\ref{SignalingIllustration}). The region between A and B is permeated by the background field $\phi_0$ which serves as a Aether-like medium for the superluminal $\varphi$ modes. A is equipped with an encoder capable of generating semi-classical signals according to the input information. B receives the signal and extract its information via a decoder.
Since we are focusing on semi-classical superluminal communication, the natural choice for the signal is a coherent state \cite{Klauder:1960kt,Glauber:1963tx},
\begin{equation}
|z\rangle=\exp\left(-\frac{1}{2}\int_{\vec{k}}|z_{\vec{k}}|^2\right)\exp\left(\int_{\vec{k}}z_{\vec{k}} a_{\vec{k}}^\dagger\right)|0\rangle~,
\end{equation}
where $z_{\vec{k}}$ is the eigenvalue of the annihilation operator, $a_{\vec{k}}|z\rangle=z_{\vec{k}}|z\rangle$. 
%In the rest frame of the background, the mode function is of the form
%\begin{equation}
%v_k(t)=\frac{c_se^{-i\omega_k t}}{\sqrt{2\omega_k}}~,~\text{with }\omega_k=c_s |\vec{k}|~.
%\end{equation}
The information content of the signal is encoded in the Fourier coefficients $\{z_{\vec{k}}\}$. Generically for point-to-point communications, the signal takes the form of a wave packet with IR cutoff scales $L_x,L_y$ and $L_z$. In addition, to maximize the efficiency of communication, the wave packet should be traveling roughly in the $+x$ direction. Hence the Fourier components along the transverse directions should be turned down, with an uncertainty of $\Delta k_y\sim \frac{1}{L_y},\Delta k_z\sim \frac{1}{L_z}$. In other words, we impose
\begin{equation}
z_{\vec{k}}=0~\text{ for }|k_y|>\frac{2\pi}{L_y}\text{ or }|k_z|>\frac{2\pi}{L_z}~.
\end{equation}
%If the signal freely propagates in the medium, the transverse size should be much larger than that of the longitudinal direction, so as to prevent the wave packet to spread out transversely\footnote{In the case with a wave guide, there is no need to keep $L_y,L_z\gg L_x$, since there is an energy gap between the transverse excitation and the ground mode.}.
By this choice, $z_{\vec{k}}$ only has support for $\vec{k}$ lying on the $+x$ direction, then the evolution of the signal is simplified to a shape-preserving translation, $i.e.$, $\varphi=\varphi(x-c_s t)$. In this way we have reduced the communication model to a one-dimensional problem while maintaining its maximal ability to transfer information. We define the expectation value of the area density of energy by
\begin{equation}
E[\hat{z}]\equiv\frac{\langle z|H^{(2)}|z\rangle}{L_y L_z}=\int\frac{dk}{2\pi}\omega_k |\hat{z}_k|^2~,\label{EnergyAreaDensity}
\end{equation}
where $\hat{z}_k\equiv\frac{z_k}{L_y L_z}$ is the reduced Fourier moment. In order to emit the signal, we turn on a weak coupling between the k-essence perturbation $\varphi$ and the visible SM sector via a shift-symmetric coupling $J^\mu_{\text{SM}}\partial_\mu\varphi$. The source $J^\mu_{\text{SM}}$ is turned on for $0<t<\frac{L_x}{c_s}$. In the free-theory limit, this can be mimicked by imposing a periodic boundary condition between the sound-cones originating from the source, leading to momentum quantization $k_n=\frac{2\pi n}{L_x}$. Notice that this is only possible in the free theory limit. In the presence of interactions, there is no automatic boundary condition inherited from the source function and the wave packet will in general spread out due to its self-interactions.

With all the preparations complete, now we can apply the constraint (\ref{ImportantConstraint}) on the field gradient to obtain our bit rate bound. In the semi-classical limit, the information content of the communication channel where the messages are sent according to a given probability distribution functional $P[\hat{z}]$ can be characterized by the Shannon entropy
\begin{equation}
S[P]=-\int\mathcal{D}^2\hat{z}P[\hat{z}]\ln P[\hat{z}]~,~\text{ with }\int\mathcal{D}^2\hat{z}P[\hat{z}]=1~.
\end{equation}
The integration measure $\mathcal{D}^2\hat{z} \propto \prod_{k>0}d\hat{z}_k d\hat{z}_k^*$ must be taken with a grain of salt. The fact that the coherent state basis is over-complete \cite{Klauder:1960kt} leads to unphysical contributions to the Shannon entropy. These contributions must be gauged away by truncating the phase space integral near the origin. Equivalently, one can start with the von Neumann entropy $S_{\text{V.N.}}[P]=-\mathrm{Tr} \rho\ln \rho$ of a density matrix $\rho[P]=\int\mathcal{D}^2\hat{z}P[\hat{z}]|z\rangle\langle z|$ describing a statistical mixture of coded signals.

We can impose three conceivable constraints. The most straightforward one from (\ref{ImportantConstraint}) is a local constraint imposed at every spacetime points throughout the interior of the wave packet,
\begin{equation}
\text{(LC): }\left|\langle z|\dot\varphi(t,x)|z\rangle\right|\lesssim \frac{\dot\phi_0}{c_s^2} ~\text{ for } c_s t-L_x<x<c_s t~.
\end{equation}
This is too restrictive since (LC) violations may not endanger the global information content of the signal. The second is a global constraint that smears (LC) over the wave packet:
\begin{equation}
\text{(GC): }E[\hat{z}]\lesssim\varepsilon_{\text{max}} L_x~,
\end{equation}
where $E[\hat{z}]$ is given by (\ref{EnergyAreaDensity}) and $\varepsilon_{\text{max}}\sim \frac{\dot\phi_0^2}{c_s^6}$ is the maximal energy density. The third is an average global constraint:
\begin{equation}
\text{(AGC): }\int \mathcal{D}^2\hat{z}P[\hat{z}] E[\hat{z}]\lesssim\varepsilon_{\text{max}} L_x~.
\end{equation}
Namely, we do not require (GC) to be satisfied by all signals generated from $P[\hat{z}]$, but only by its average output. This is also more suitable compared to (GC), in the sense that the failure of selective signals does not affect the efficiency of the whole coding method $P[\hat{z}]$. The violation of (AGC) would imply non-linearity so large that signals cannot be prepared or even defined within the theory (\ref{expandedLagrangian}), $i.e.$, an IR breakdown of the EFT. It is useful to notice the inclusion relation between the constraints,
\begin{equation}
(\text{LC})\subset(\text{GC})\subseteq(\text{AGC})~.
\end{equation}

Now it is straightforward to bound the Shannon entropy by maximizing it under (AGC). Introducing two Lagrange multipliers $\alpha,\beta$, we seek the stationary point by
\begin{equation}
0=\delta\left(S[P]-(\alpha-1)\int \mathcal{D}^2\hat{z} P-\beta\int\mathcal{D}^2\hat{z}PE\right)~.
\end{equation}
The optimal probability distribution then takes a familiar form, $P_*=e^{-\alpha-\beta E}$. The Lagrange multipliers, like in thermodynamics, are solved from the constraints. The maximal entropy is then
\begin{equation}
S_*=\frac{L_x \Lambda}{2\pi c_s}\ln\left(1+\frac{1}{\gamma }e^{-\gamma}\right)+\frac{2\dot\phi_0^2L_x L_y L_z}{c_s^6\Lambda}\gamma~,
\end{equation}
where $\gamma$ is the solution to
\begin{equation}
\frac{1}{\gamma^2}\int_{0}^{\gamma}dx\frac{1+x}{1+x e^x}=\frac{2\pi \dot\phi_0^2 L_y L_z}{c_s^5 \Lambda^2}~.\label{gammaEquation}
\end{equation}
Thus the (AGC) bound for bit rate is given by
\begin{equation}
R_*=\frac{c_s S_*}{L_x}\equiv\Lambda\mathcal{R}_*(c_s,\dot\phi_0^2 L_y L_z/\Lambda^2)~.
\end{equation}
This bit rate bound simplifies considerably in two limits. In the small-cross section limit where $\frac{\dot\phi_0^2 L_y L_z}{\Lambda^2}\ll \frac{c_s^5}{2\pi}$, the bit rate bound is independent of $\Lambda$,
\begin{equation}
R_*\approx\left(\frac{2.64}{\pi}\dot\phi_0^2 L_y L_z\right)^{1/2} c_s^{-5/2}~.\label{lowTempLimit}
\end{equation}
On the other hand, if the communication cross section is large, $i.e.$, $\frac{\dot\phi_0^2 L_y L_z}{\Lambda^2}\gg \frac{c_s^5}{2\pi}$, the EFT cutoff $\Lambda$ explicitly appears in the bit rate bound,
\begin{equation}
R_*\approx\frac{\Lambda}{2\pi}\ln\frac{2\pi \dot\phi_0^2 L_y L_z}{c_s^5 \Lambda^2}~.
\end{equation}
\begin{figure}[h!]
	\centering
	\includegraphics[width=7.8cm]{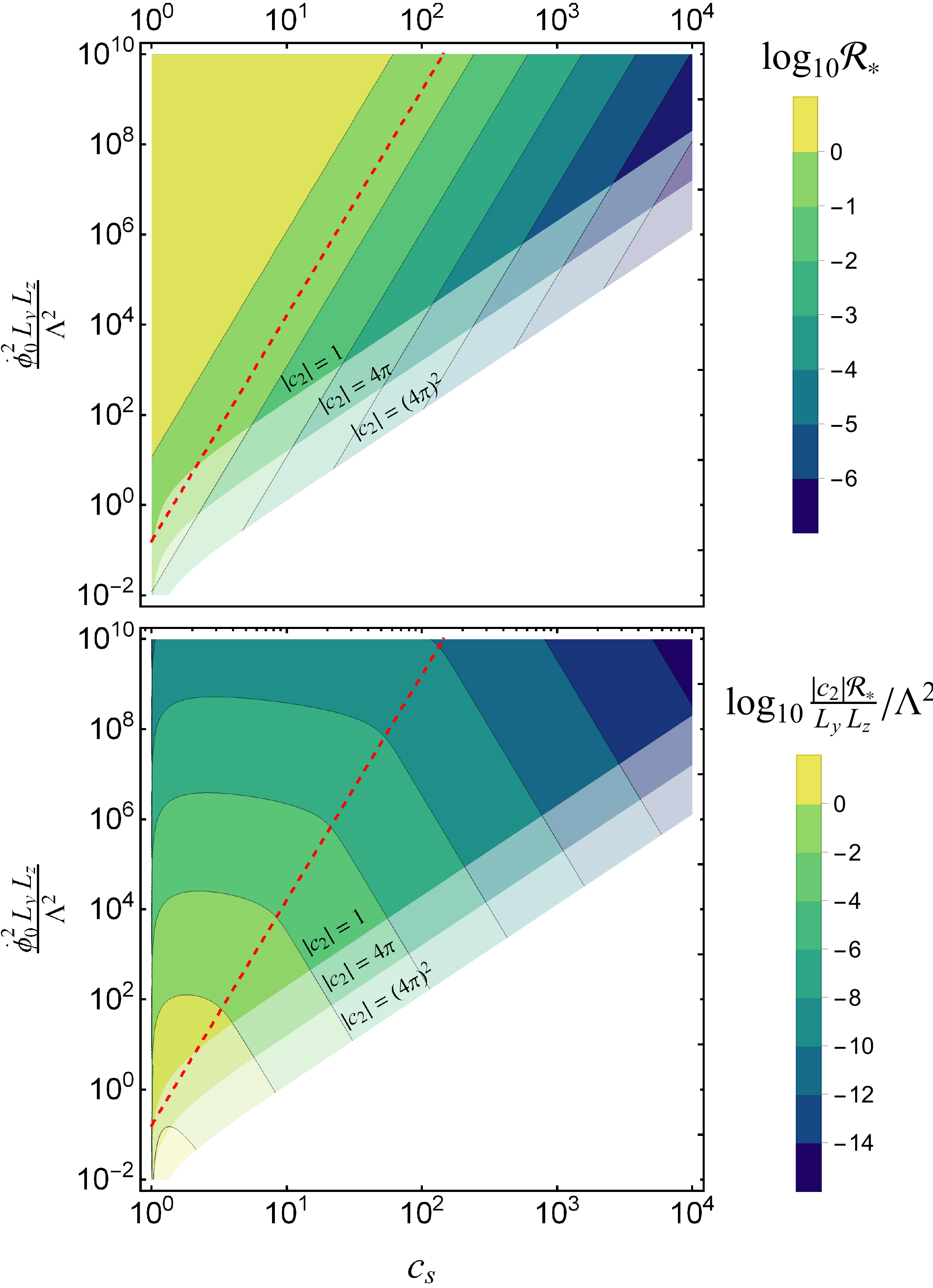}
	\caption{Dimensionless bit rate bound $\mathcal{R}_*$ and its area density. The red dashed line is given by $\frac{\dot\phi_0^2 L_y L_z}{\Lambda^2}=\frac{c_s^5}{2\pi}$. The region above this line is the logarithmic high-temperature case while that (much) below the line is the power-law low-temperature case. The lower regions are excluded due to EFT validity for given choices of $|c_2|$.}\label{BitRate2d}
\end{figure}
The full dependence of the dimensionless bit rate $\mathcal{R}_*$ as well as its area density is shown in FIG.~\ref{BitRate2d}. Obviously it is monotonically decreasing with $c_s$. As a result, increasing the propagation speed indefinitely \textit{cannot} always lead to an increasing bit rate, since its upper bound drops with $c_s$. Moreover, although mathematically $\mathcal{R}_*$ is not bounded from above due to its logarithmic growth with the cross section area $L_y L_z$, the bit rate area density is, however, decreasing with $L_y L_z$. Therefore, the bit rate area density is universally bounded by
\begin{equation}
\frac{R_*}{L_y L_z}\lesssim\left(\frac{2.64}{\pi}\right)^{1/2}\dot\phi_0\Lambda c_s^{-7/2}\approx 1.30|c_2|^{-1/2}c_s^{-7/2}\Lambda^3~,\label{BitRateAreaDensityBound}
\end{equation}
where we have set a minimal area $L_y L_z\sim c_s^2\Lambda^{-2}$ consistent with the EFT in (\ref{lowTempLimit}). Then (\ref{BitRateAreaDensityBound}) gives a maximal ability to compress the signal in the transverse direction\footnote{In free space, too much restriction in the transverse direction generically leads to a fast spreading wave packet. This can be prevented, for example, by using a narrow wave guide, where transverse excitations are energetically too expensive to be excited.}. Namely, bit rate per unit area is still bounded.

We point out that our treatment resembles the description of a one-dimensional photon gas confined in a box with a modified dispersion relation. The small-cross-section case corresponds to its low-temperature behavior, where high-frequency modes are automatically suppressed by the Boltzmann factor and the cutoff frequency $\Lambda$ is never reached. The large-cross-section limit corresponds to the high temperature case, where modes near the cutoff can also be excited with abundance. But this time, the validity of EFT puts a sharp cutoff to the available modes, and the final result is dependent on $\Lambda$, a reminiscent of the UV catastrophe in classical physics. %In statistical mechanics, there is no explicit cutoff on the energy of a photon, yet the high-energy occupation number is still Boltzmann-suppressed, therefore leading to a convergent energy integral $\sum P_i E_i=\bar{E}$. This is not always so in history. Before 1900, there is the infamous ultraviolet catastrophe, where classical argument gives $P_i\propto E_i^2$ and a divergent energy integral. Of course, the dilemma was solved by Planck's famous proposal of energy quanta proposed by Planck, the first raindrop of a storm of physics revolution. Without realizing the quantum nature of the coherent state and replacing the phase space integral in (\ref{adhocModification}), we would have run into the same ultraviolet catastrophe even in the small-cross-section case, with the only difference being that we do have a physical cutoff here. This is indeed what happens in the large-cross-section case, or equivalently, in the high-temperature limit of photon gas, namely, classical physics re-surges when occupation numbers are huge or temperature is high.

There are, of course, two important distinctions. The first one being conceptual, is that we are not actually sending a thermal mixture of states confined in a one-dimensional box. We have precise control over the signal, which is in a \textit{pure} state. The entropy resides in the coding method we use. The maximal entropy under (AGC) is reached by coding the messages following a Boltzmann distribution. Of course, one does not have to code the messages this way. Freedom is given to the sender A such that any probabilistic distribution $P[\hat{z}]$ is allowed, although none of them can exceed the efficiency provided by $P_*[\hat{z}]$.

The second one being technical, is that we are counting coherent states of the k-essence field rather than their energy eigenstates (when confined to a box). The coherent states are almost classical, whereas the energy eigenstates are non-classical. When the non-linear interaction of k-essence is turned down, either choice is fine. In fact, a simple calculation shows that they give the same Shannon entropy $S_*$ up to an $\mathcal{O}(1)$ numerical factor.

%\section{Conclusion}
{\it Conclusion.} --
In this {\it Letter} we have investigated superluminal communication in a family of positivity-violating k-essence theories. After setting up a conceptual superluminal communication protocol, by the requirement of subdominant non-linearity, we have derived a bit rate bound and a universal bit rate area density bound for all superluminal k-essence models with a natural choice of Wilson coefficients. Notice that is this work we focus exclusively on the bit rate bound intrinsic to the k-essence theory itself. The phenomenological implications of coupling it to the SM and a detailed analysis of the signal evolution are left for future works (in progress). Finally, we remark that k-essence is not the only superluminal communication candidate. It is interesting to consider whether similar bit rate bounds exist for other methods \cite{Gubser:2011mp,Ramos:2019one}.

%\section*{Acknowledgments}
{\it Acknowledgments.} --
We thank Ali Akil, Lingfeng Li, Kun-Feng Lyu and Henry Tye for helpful discussions and suggestions. This work was supported in part by the NSFC Excellent Young Scientist (EYS) Scheme (Hong Kong and Macau) Grant No.~12022516.
\\

\bibliography{reference}

%merlin.mbs apsrev4-1.bst 2010-07-25 4.21a (PWD, AO, DPC) hacked
%Control: key (0)
%Control: author (72) initials jnrlst
%Control: editor formatted (1) identically to author
%Control: production of article title (-1) disabled
%Control: page (0) single
%Control: year (1) truncated
%Control: production of eprint (0) enabled
\begin{thebibliography}{40}%
\makeatletter
\providecommand \@ifxundefined [1]{%
 \@ifx{#1\undefined}
}%
\providecommand \@ifnum [1]{%
 \ifnum #1\expandafter \@firstoftwo
 \else \expandafter \@secondoftwo
 \fi
}%
\providecommand \@ifx [1]{%
 \ifx #1\expandafter \@firstoftwo
 \else \expandafter \@secondoftwo
 \fi
}%
\providecommand \natexlab [1]{#1}%
\providecommand \enquote  [1]{``#1''}%
\providecommand \bibnamefont  [1]{#1}%
\providecommand \bibfnamefont [1]{#1}%
\providecommand \citenamefont [1]{#1}%
\providecommand \href@noop [0]{\@secondoftwo}%
\providecommand \href [0]{\begingroup \@sanitize@url \@href}%
\providecommand \@href[1]{\@@startlink{#1}\@@href}%
\providecommand \@@href[1]{\endgroup#1\@@endlink}%
\providecommand \@sanitize@url [0]{\catcode `\\12\catcode `\$12\catcode
  `\&12\catcode `\#12\catcode `\^12\catcode `\_12\catcode `\%12\relax}%
\providecommand \@@startlink[1]{}%
\providecommand \@@endlink[0]{}%
\providecommand \url  [0]{\begingroup\@sanitize@url \@url }%
\providecommand \@url [1]{\endgroup\@href {#1}{\urlprefix }}%
\providecommand \urlprefix  [0]{URL }%
\providecommand \Eprint [0]{\href }%
\providecommand \doibase [0]{http://dx.doi.org/}%
\providecommand \selectlanguage [0]{\@gobble}%
\providecommand \bibinfo  [0]{\@secondoftwo}%
\providecommand \bibfield  [0]{\@secondoftwo}%
\providecommand \translation [1]{[#1]}%
\providecommand \BibitemOpen [0]{}%
\providecommand \bibitemStop [0]{}%
\providecommand \bibitemNoStop [0]{.\EOS\space}%
\providecommand \EOS [0]{\spacefactor3000\relax}%
\providecommand \BibitemShut  [1]{\csname bibitem#1\endcsname}%
\let\auto@bib@innerbib\@empty
%</preamble>
\bibitem [{\citenamefont {Einstein}(1905)}]{Einstein:1905ve}%
  \BibitemOpen
  \bibfield  {author} {\bibinfo {author} {\bibfnamefont {A.}~\bibnamefont
  {Einstein}},\ }\href {\doibase 10.1002/andp.200590006} {\bibfield  {journal}
  {\bibinfo  {journal} {Annalen Phys.}\ }\textbf {\bibinfo {volume} {17}},\
  \bibinfo {pages} {891} (\bibinfo {year} {1905})}\BibitemShut {NoStop}%
\bibitem [{\citenamefont {Einstein}(1907)}]{Einstein:1907iag}%
  \BibitemOpen
  \bibfield  {author} {\bibinfo {author} {\bibfnamefont {A.}~\bibnamefont
  {Einstein}},\ }\href {\doibase 10.1002/andp.19073280713} {\bibfield
  {journal} {\bibinfo  {journal} {Annalen Phys.}\ }\textbf {\bibinfo {volume}
  {23}},\ \bibinfo {pages} {371} (\bibinfo {year} {1907})}\BibitemShut
  {NoStop}%
\bibitem [{\citenamefont {Brillouin}(1960)}]{Brillouin:1960tos}%
  \BibitemOpen
  \bibfield  {author} {\bibinfo {author} {\bibfnamefont {L.}~\bibnamefont
  {Brillouin}},\ }\href@noop {} {\emph {\bibinfo {title} {{Wave propagation and
  group velocity}}}},\ Vol.~\bibinfo {volume} {8}\ (\bibinfo  {publisher}
  {Academic Press},\ \bibinfo {address} {New York, London},\ \bibinfo {year}
  {1960})\BibitemShut {NoStop}%
\bibitem [{\citenamefont {Fox}\ \emph {et~al.}(1969)\citenamefont {Fox},
  \citenamefont {Kuper},\ and\ \citenamefont {Lipson}}]{Fox:1969us}%
  \BibitemOpen
  \bibfield  {author} {\bibinfo {author} {\bibfnamefont {R.}~\bibnamefont
  {Fox}}, \bibinfo {author} {\bibfnamefont {C.}~\bibnamefont {Kuper}}, \ and\
  \bibinfo {author} {\bibfnamefont {S.}~\bibnamefont {Lipson}},\ }\href
  {\doibase 10.1038/223597a0} {\bibfield  {journal} {\bibinfo  {journal}
  {Nature}\ }\textbf {\bibinfo {volume} {223}},\ \bibinfo {pages} {597}
  (\bibinfo {year} {1969})}\BibitemShut {NoStop}%
\bibitem [{\citenamefont {{Rees}}(1966)}]{1966Natur.211..468R}%
  \BibitemOpen
  \bibfield  {author} {\bibinfo {author} {\bibfnamefont {M.~J.}\ \bibnamefont
  {{Rees}}},\ }\href {\doibase 10.1038/211468a0} {\bibfield  {journal}
  {\bibinfo  {journal} {\nat}\ }\textbf {\bibinfo {volume} {211}},\ \bibinfo
  {pages} {468} (\bibinfo {year} {1966})}\BibitemShut {NoStop}%
\bibitem [{\citenamefont {Davis}\ and\ \citenamefont
  {Lineweaver}(2004)}]{Davis:2003ad}%
  \BibitemOpen
  \bibfield  {author} {\bibinfo {author} {\bibfnamefont {T.~M.}\ \bibnamefont
  {Davis}}\ and\ \bibinfo {author} {\bibfnamefont {C.~H.}\ \bibnamefont
  {Lineweaver}},\ }\href {\doibase 10.1071/AS03040} {\bibfield  {journal}
  {\bibinfo  {journal} {Publ. Astron. Soc. Austral.}\ }\textbf {\bibinfo
  {volume} {21}},\ \bibinfo {pages} {97} (\bibinfo {year} {2004})},\ \Eprint
  {http://arxiv.org/abs/astro-ph/0310808} {arXiv:astro-ph/0310808} \BibitemShut
  {NoStop}%
\bibitem [{\citenamefont {{Fierz}}(1939)}]{1939AcHPh..12....3F}%
  \BibitemOpen
  \bibfield  {author} {\bibinfo {author} {\bibfnamefont {M.}~\bibnamefont
  {{Fierz}}},\ }\href@noop {} {\bibfield  {journal} {\bibinfo  {journal}
  {Helvetica Physica Acta}\ }\textbf {\bibinfo {volume} {12}},\ \bibinfo
  {pages} {3} (\bibinfo {year} {1939})}\BibitemShut {NoStop}%
\bibitem [{\citenamefont {Pauli}(1940)}]{Pauli:1940zz}%
  \BibitemOpen
  \bibfield  {author} {\bibinfo {author} {\bibfnamefont {W.}~\bibnamefont
  {Pauli}},\ }\href {\doibase 10.1103/PhysRev.58.716} {\bibfield  {journal}
  {\bibinfo  {journal} {Phys. Rev.}\ }\textbf {\bibinfo {volume} {58}},\
  \bibinfo {pages} {716} (\bibinfo {year} {1940})}\BibitemShut {NoStop}%
\bibitem [{\citenamefont {Luders}(1954)}]{Luders:1954zz}%
  \BibitemOpen
  \bibfield  {author} {\bibinfo {author} {\bibfnamefont {G.}~\bibnamefont
  {Luders}},\ }\href@noop {} {\bibfield  {journal} {\bibinfo  {journal} {Kong.
  Dan. Vid. Sel. Mat. Fys. Med.}\ }\textbf {\bibinfo {volume} {28N5}},\
  \bibinfo {pages} {1} (\bibinfo {year} {1954})}\BibitemShut {NoStop}%
\bibitem [{\citenamefont {Weinberg}(1964)}]{Weinberg:1964ew}%
  \BibitemOpen
  \bibfield  {author} {\bibinfo {author} {\bibfnamefont {S.}~\bibnamefont
  {Weinberg}},\ }\href {\doibase 10.1103/PhysRev.135.B1049} {\bibfield
  {journal} {\bibinfo  {journal} {Phys. Rev.}\ }\textbf {\bibinfo {volume}
  {135}},\ \bibinfo {pages} {B1049} (\bibinfo {year} {1964})}\BibitemShut
  {NoStop}%
\bibitem [{\citenamefont {Coleman}\ and\ \citenamefont
  {Mandula}(1967)}]{Coleman:1967ad}%
  \BibitemOpen
  \bibfield  {author} {\bibinfo {author} {\bibfnamefont {S.~R.}\ \bibnamefont
  {Coleman}}\ and\ \bibinfo {author} {\bibfnamefont {J.}~\bibnamefont
  {Mandula}},\ }\href {\doibase 10.1103/PhysRev.159.1251} {\bibfield  {journal}
  {\bibinfo  {journal} {Phys. Rev.}\ }\textbf {\bibinfo {volume} {159}},\
  \bibinfo {pages} {1251} (\bibinfo {year} {1967})}\BibitemShut {NoStop}%
\bibitem [{\citenamefont {Weinberg}\ and\ \citenamefont
  {Witten}(1980)}]{Weinberg:1980kq}%
  \BibitemOpen
  \bibfield  {author} {\bibinfo {author} {\bibfnamefont {S.}~\bibnamefont
  {Weinberg}}\ and\ \bibinfo {author} {\bibfnamefont {E.}~\bibnamefont
  {Witten}},\ }\href {\doibase 10.1016/0370-2693(80)90212-9} {\bibfield
  {journal} {\bibinfo  {journal} {Phys. Lett. B}\ }\textbf {\bibinfo {volume}
  {96}},\ \bibinfo {pages} {59} (\bibinfo {year} {1980})}\BibitemShut {NoStop}%
\bibitem [{\citenamefont {Colladay}\ and\ \citenamefont
  {Kostelecky}(1998)}]{Colladay:1998fq}%
  \BibitemOpen
  \bibfield  {author} {\bibinfo {author} {\bibfnamefont {D.}~\bibnamefont
  {Colladay}}\ and\ \bibinfo {author} {\bibfnamefont {V.}~\bibnamefont
  {Kostelecky}},\ }\href {\doibase 10.1103/PhysRevD.58.116002} {\bibfield
  {journal} {\bibinfo  {journal} {Phys. Rev. D}\ }\textbf {\bibinfo {volume}
  {58}},\ \bibinfo {pages} {116002} (\bibinfo {year} {1998})},\ \Eprint
  {http://arxiv.org/abs/hep-ph/9809521} {arXiv:hep-ph/9809521} \BibitemShut
  {NoStop}%
\bibitem [{\citenamefont {Coleman}\ and\ \citenamefont
  {Glashow}(1999)}]{Coleman:1998ti}%
  \BibitemOpen
  \bibfield  {author} {\bibinfo {author} {\bibfnamefont {S.~R.}\ \bibnamefont
  {Coleman}}\ and\ \bibinfo {author} {\bibfnamefont {S.~L.}\ \bibnamefont
  {Glashow}},\ }\href {\doibase 10.1103/PhysRevD.59.116008} {\bibfield
  {journal} {\bibinfo  {journal} {Phys. Rev. D}\ }\textbf {\bibinfo {volume}
  {59}},\ \bibinfo {pages} {116008} (\bibinfo {year} {1999})},\ \Eprint
  {http://arxiv.org/abs/hep-ph/9812418} {arXiv:hep-ph/9812418} \BibitemShut
  {NoStop}%
\bibitem [{\citenamefont {Kostelecky}\ and\ \citenamefont
  {Russell}(2008)}]{Kostelecky:2008ts}%
  \BibitemOpen
  \bibfield  {author} {\bibinfo {author} {\bibfnamefont {V.}~\bibnamefont
  {Kostelecky}}\ and\ \bibinfo {author} {\bibfnamefont {N.}~\bibnamefont
  {Russell}},\ }\href@noop {} {\  (\bibinfo {year} {2008})},\ \Eprint
  {http://arxiv.org/abs/0801.0287} {arXiv:0801.0287 [hep-ph]} \BibitemShut
  {NoStop}%
\bibitem [{\citenamefont {Cohen}\ and\ \citenamefont
  {Glashow}(2011)}]{Cohen:2011hx}%
  \BibitemOpen
  \bibfield  {author} {\bibinfo {author} {\bibfnamefont {A.~G.}\ \bibnamefont
  {Cohen}}\ and\ \bibinfo {author} {\bibfnamefont {S.~L.}\ \bibnamefont
  {Glashow}},\ }\href {\doibase 10.1103/PhysRevLett.107.181803} {\bibfield
  {journal} {\bibinfo  {journal} {Phys. Rev. Lett.}\ }\textbf {\bibinfo
  {volume} {107}},\ \bibinfo {pages} {181803} (\bibinfo {year} {2011})},\
  \Eprint {http://arxiv.org/abs/1109.6562} {arXiv:1109.6562 [hep-ph]}
  \BibitemShut {NoStop}%
\bibitem [{\citenamefont {Liberati}(2013)}]{Liberati:2013xla}%
  \BibitemOpen
  \bibfield  {author} {\bibinfo {author} {\bibfnamefont {S.}~\bibnamefont
  {Liberati}},\ }\href {\doibase 10.1088/0264-9381/30/13/133001} {\bibfield
  {journal} {\bibinfo  {journal} {Class. Quant. Grav.}\ }\textbf {\bibinfo
  {volume} {30}},\ \bibinfo {pages} {133001} (\bibinfo {year} {2013})},\
  \Eprint {http://arxiv.org/abs/1304.5795} {arXiv:1304.5795 [gr-qc]}
  \BibitemShut {NoStop}%
\bibitem [{\citenamefont {Armendariz-Picon}\ \emph {et~al.}(1999)\citenamefont
  {Armendariz-Picon}, \citenamefont {Damour},\ and\ \citenamefont
  {Mukhanov}}]{ArmendarizPicon:1999rj}%
  \BibitemOpen
  \bibfield  {author} {\bibinfo {author} {\bibfnamefont {C.}~\bibnamefont
  {Armendariz-Picon}}, \bibinfo {author} {\bibfnamefont {T.}~\bibnamefont
  {Damour}}, \ and\ \bibinfo {author} {\bibfnamefont {V.~F.}\ \bibnamefont
  {Mukhanov}},\ }\href {\doibase 10.1016/S0370-2693(99)00603-6} {\bibfield
  {journal} {\bibinfo  {journal} {Phys. Lett. B}\ }\textbf {\bibinfo {volume}
  {458}},\ \bibinfo {pages} {209} (\bibinfo {year} {1999})},\ \Eprint
  {http://arxiv.org/abs/hep-th/9904075} {arXiv:hep-th/9904075} \BibitemShut
  {NoStop}%
\bibitem [{\citenamefont {Garriga}\ and\ \citenamefont
  {Mukhanov}(1999)}]{Garriga:1999vw}%
  \BibitemOpen
  \bibfield  {author} {\bibinfo {author} {\bibfnamefont {J.}~\bibnamefont
  {Garriga}}\ and\ \bibinfo {author} {\bibfnamefont {V.~F.}\ \bibnamefont
  {Mukhanov}},\ }\href {\doibase 10.1016/S0370-2693(99)00602-4} {\bibfield
  {journal} {\bibinfo  {journal} {Phys. Lett. B}\ }\textbf {\bibinfo {volume}
  {458}},\ \bibinfo {pages} {219} (\bibinfo {year} {1999})},\ \Eprint
  {http://arxiv.org/abs/hep-th/9904176} {arXiv:hep-th/9904176} \BibitemShut
  {NoStop}%
\bibitem [{\citenamefont {Armendariz-Picon}\ \emph {et~al.}(2000)\citenamefont
  {Armendariz-Picon}, \citenamefont {Mukhanov},\ and\ \citenamefont
  {Steinhardt}}]{ArmendarizPicon:2000dh}%
  \BibitemOpen
  \bibfield  {author} {\bibinfo {author} {\bibfnamefont {C.}~\bibnamefont
  {Armendariz-Picon}}, \bibinfo {author} {\bibfnamefont {V.~F.}\ \bibnamefont
  {Mukhanov}}, \ and\ \bibinfo {author} {\bibfnamefont {P.~J.}\ \bibnamefont
  {Steinhardt}},\ }\href {\doibase 10.1103/PhysRevLett.85.4438} {\bibfield
  {journal} {\bibinfo  {journal} {Phys. Rev. Lett.}\ }\textbf {\bibinfo
  {volume} {85}},\ \bibinfo {pages} {4438} (\bibinfo {year} {2000})},\ \Eprint
  {http://arxiv.org/abs/astro-ph/0004134} {arXiv:astro-ph/0004134} \BibitemShut
  {NoStop}%
\bibitem [{\citenamefont {Armendariz-Picon}\ \emph {et~al.}(2001)\citenamefont
  {Armendariz-Picon}, \citenamefont {Mukhanov},\ and\ \citenamefont
  {Steinhardt}}]{ArmendarizPicon:2000ah}%
  \BibitemOpen
  \bibfield  {author} {\bibinfo {author} {\bibfnamefont {C.}~\bibnamefont
  {Armendariz-Picon}}, \bibinfo {author} {\bibfnamefont {V.~F.}\ \bibnamefont
  {Mukhanov}}, \ and\ \bibinfo {author} {\bibfnamefont {P.~J.}\ \bibnamefont
  {Steinhardt}},\ }\href {\doibase 10.1103/PhysRevD.63.103510} {\bibfield
  {journal} {\bibinfo  {journal} {Phys. Rev. D}\ }\textbf {\bibinfo {volume}
  {63}},\ \bibinfo {pages} {103510} (\bibinfo {year} {2001})},\ \Eprint
  {http://arxiv.org/abs/astro-ph/0006373} {arXiv:astro-ph/0006373} \BibitemShut
  {NoStop}%
\bibitem [{\citenamefont {Mukhanov}\ and\ \citenamefont
  {Vikman}(2006)}]{Mukhanov:2005bu}%
  \BibitemOpen
  \bibfield  {author} {\bibinfo {author} {\bibfnamefont {V.~F.}\ \bibnamefont
  {Mukhanov}}\ and\ \bibinfo {author} {\bibfnamefont {A.}~\bibnamefont
  {Vikman}},\ }\href {\doibase 10.1088/1475-7516/2006/02/004} {\bibfield
  {journal} {\bibinfo  {journal} {JCAP}\ }\textbf {\bibinfo {volume} {02}},\
  \bibinfo {pages} {004} (\bibinfo {year} {2006})},\ \Eprint
  {http://arxiv.org/abs/astro-ph/0512066} {arXiv:astro-ph/0512066} \BibitemShut
  {NoStop}%
\bibitem [{\citenamefont {Drummond}\ and\ \citenamefont
  {Hathrell}(1980)}]{Drummond:1979pp}%
  \BibitemOpen
  \bibfield  {author} {\bibinfo {author} {\bibfnamefont {I.}~\bibnamefont
  {Drummond}}\ and\ \bibinfo {author} {\bibfnamefont {S.}~\bibnamefont
  {Hathrell}},\ }\href {\doibase 10.1103/PhysRevD.22.343} {\bibfield  {journal}
  {\bibinfo  {journal} {Phys. Rev. D}\ }\textbf {\bibinfo {volume} {22}},\
  \bibinfo {pages} {343} (\bibinfo {year} {1980})}\BibitemShut {NoStop}%
\bibitem [{\citenamefont {Scharnhorst}(1990)}]{Scharnhorst:1990sr}%
  \BibitemOpen
  \bibfield  {author} {\bibinfo {author} {\bibfnamefont {K.}~\bibnamefont
  {Scharnhorst}},\ }\href {\doibase 10.1016/0370-2693(90)90997-K} {\bibfield
  {journal} {\bibinfo  {journal} {Phys. Lett. B}\ }\textbf {\bibinfo {volume}
  {236}},\ \bibinfo {pages} {354} (\bibinfo {year} {1990})},\ \bibinfo {note}
  {[Erratum: Phys.Lett.B 787, 204--204 (2018)]}\BibitemShut {NoStop}%
\bibitem [{\citenamefont {Barton}(1990)}]{Barton:1989dq}%
  \BibitemOpen
  \bibfield  {author} {\bibinfo {author} {\bibfnamefont {G.}~\bibnamefont
  {Barton}},\ }\href {\doibase 10.1016/0370-2693(90)91224-Y} {\bibfield
  {journal} {\bibinfo  {journal} {Phys. Lett. B}\ }\textbf {\bibinfo {volume}
  {237}},\ \bibinfo {pages} {559} (\bibinfo {year} {1990})}\BibitemShut
  {NoStop}%
\bibitem [{\citenamefont {Jacobson}\ and\ \citenamefont
  {Mattingly}(2004)}]{Jacobson:2004ts}%
  \BibitemOpen
  \bibfield  {author} {\bibinfo {author} {\bibfnamefont {T.}~\bibnamefont
  {Jacobson}}\ and\ \bibinfo {author} {\bibfnamefont {D.}~\bibnamefont
  {Mattingly}},\ }\href {\doibase 10.1103/PhysRevD.70.024003} {\bibfield
  {journal} {\bibinfo  {journal} {Phys. Rev. D}\ }\textbf {\bibinfo {volume}
  {70}},\ \bibinfo {pages} {024003} (\bibinfo {year} {2004})},\ \Eprint
  {http://arxiv.org/abs/gr-qc/0402005} {arXiv:gr-qc/0402005} \BibitemShut
  {NoStop}%
\bibitem [{\citenamefont {Adams}\ \emph {et~al.}(2006)\citenamefont {Adams},
  \citenamefont {Arkani-Hamed}, \citenamefont {Dubovsky}, \citenamefont
  {Nicolis},\ and\ \citenamefont {Rattazzi}}]{Adams:2006sv}%
  \BibitemOpen
  \bibfield  {author} {\bibinfo {author} {\bibfnamefont {A.}~\bibnamefont
  {Adams}}, \bibinfo {author} {\bibfnamefont {N.}~\bibnamefont {Arkani-Hamed}},
  \bibinfo {author} {\bibfnamefont {S.}~\bibnamefont {Dubovsky}}, \bibinfo
  {author} {\bibfnamefont {A.}~\bibnamefont {Nicolis}}, \ and\ \bibinfo
  {author} {\bibfnamefont {R.}~\bibnamefont {Rattazzi}},\ }\href {\doibase
  10.1088/1126-6708/2006/10/014} {\bibfield  {journal} {\bibinfo  {journal}
  {JHEP}\ }\textbf {\bibinfo {volume} {10}},\ \bibinfo {pages} {014} (\bibinfo
  {year} {2006})},\ \Eprint {http://arxiv.org/abs/hep-th/0602178}
  {arXiv:hep-th/0602178} \BibitemShut {NoStop}%
\bibitem [{\citenamefont {Shore}(2007)}]{Shore:2007um}%
  \BibitemOpen
  \bibfield  {author} {\bibinfo {author} {\bibfnamefont {G.}~\bibnamefont
  {Shore}},\ }\href {\doibase 10.1016/j.nuclphysb.2007.03.034} {\bibfield
  {journal} {\bibinfo  {journal} {Nucl. Phys. B}\ }\textbf {\bibinfo {volume}
  {778}},\ \bibinfo {pages} {219} (\bibinfo {year} {2007})},\ \Eprint
  {http://arxiv.org/abs/hep-th/0701185} {arXiv:hep-th/0701185} \BibitemShut
  {NoStop}%
\bibitem [{\citenamefont {Armendariz-Picon}\ and\ \citenamefont
  {Lim}(2005)}]{ArmendarizPicon:2005nz}%
  \BibitemOpen
  \bibfield  {author} {\bibinfo {author} {\bibfnamefont {C.}~\bibnamefont
  {Armendariz-Picon}}\ and\ \bibinfo {author} {\bibfnamefont {E.~A.}\
  \bibnamefont {Lim}},\ }\href {\doibase 10.1088/1475-7516/2005/08/007}
  {\bibfield  {journal} {\bibinfo  {journal} {JCAP}\ }\textbf {\bibinfo
  {volume} {08}},\ \bibinfo {pages} {007} (\bibinfo {year} {2005})},\ \Eprint
  {http://arxiv.org/abs/astro-ph/0505207} {arXiv:astro-ph/0505207} \BibitemShut
  {NoStop}%
\bibitem [{\citenamefont {Bruneton}(2007)}]{Bruneton:2006gf}%
  \BibitemOpen
  \bibfield  {author} {\bibinfo {author} {\bibfnamefont {J.-P.}\ \bibnamefont
  {Bruneton}},\ }\href {\doibase 10.1103/PhysRevD.75.085013} {\bibfield
  {journal} {\bibinfo  {journal} {Phys. Rev. D}\ }\textbf {\bibinfo {volume}
  {75}},\ \bibinfo {pages} {085013} (\bibinfo {year} {2007})},\ \Eprint
  {http://arxiv.org/abs/gr-qc/0607055} {arXiv:gr-qc/0607055} \BibitemShut
  {NoStop}%
\bibitem [{\citenamefont {Babichev}\ \emph {et~al.}(2008)\citenamefont
  {Babichev}, \citenamefont {Mukhanov},\ and\ \citenamefont
  {Vikman}}]{Babichev:2007dw}%
  \BibitemOpen
  \bibfield  {author} {\bibinfo {author} {\bibfnamefont {E.}~\bibnamefont
  {Babichev}}, \bibinfo {author} {\bibfnamefont {V.}~\bibnamefont {Mukhanov}},
  \ and\ \bibinfo {author} {\bibfnamefont {A.}~\bibnamefont {Vikman}},\ }\href
  {\doibase 10.1088/1126-6708/2008/02/101} {\bibfield  {journal} {\bibinfo
  {journal} {JHEP}\ }\textbf {\bibinfo {volume} {02}},\ \bibinfo {pages} {101}
  (\bibinfo {year} {2008})},\ \Eprint {http://arxiv.org/abs/0708.0561}
  {arXiv:0708.0561 [hep-th]} \BibitemShut {NoStop}%
\bibitem [{\citenamefont {Wald}(1984)}]{Wald:1984rg}%
  \BibitemOpen
  \bibfield  {author} {\bibinfo {author} {\bibfnamefont {R.~M.}\ \bibnamefont
  {Wald}},\ }\href {\doibase 10.7208/chicago/9780226870373.001.0001} {\emph
  {\bibinfo {title} {{General Relativity}}}}\ (\bibinfo  {publisher} {Chicago
  Univ. Pr.},\ \bibinfo {address} {Chicago, USA},\ \bibinfo {year}
  {1984})\BibitemShut {NoStop}%
\bibitem [{\citenamefont {Barcelo}\ \emph {et~al.}(2005)\citenamefont
  {Barcelo}, \citenamefont {Liberati},\ and\ \citenamefont
  {Visser}}]{Barcelo:2005fc}%
  \BibitemOpen
  \bibfield  {author} {\bibinfo {author} {\bibfnamefont {C.}~\bibnamefont
  {Barcelo}}, \bibinfo {author} {\bibfnamefont {S.}~\bibnamefont {Liberati}}, \
  and\ \bibinfo {author} {\bibfnamefont {M.}~\bibnamefont {Visser}},\ }\href
  {\doibase 10.12942/lrr-2005-12} {\bibfield  {journal} {\bibinfo  {journal}
  {Living Rev. Rel.}\ }\textbf {\bibinfo {volume} {8}},\ \bibinfo {pages} {12}
  (\bibinfo {year} {2005})},\ \Eprint {http://arxiv.org/abs/gr-qc/0505065}
  {arXiv:gr-qc/0505065} \BibitemShut {NoStop}%
\bibitem [{\citenamefont {Pham}\ and\ \citenamefont
  {Truong}(1985)}]{Pham:1985cr}%
  \BibitemOpen
  \bibfield  {author} {\bibinfo {author} {\bibfnamefont {T.}~\bibnamefont
  {Pham}}\ and\ \bibinfo {author} {\bibfnamefont {T.~N.}\ \bibnamefont
  {Truong}},\ }\href {\doibase 10.1103/PhysRevD.31.3027} {\bibfield  {journal}
  {\bibinfo  {journal} {Phys. Rev. D}\ }\textbf {\bibinfo {volume} {31}},\
  \bibinfo {pages} {3027} (\bibinfo {year} {1985})}\BibitemShut {NoStop}%
\bibitem [{\citenamefont {Nicolis}\ \emph {et~al.}(2010)\citenamefont
  {Nicolis}, \citenamefont {Rattazzi},\ and\ \citenamefont
  {Trincherini}}]{Nicolis:2009qm}%
  \BibitemOpen
  \bibfield  {author} {\bibinfo {author} {\bibfnamefont {A.}~\bibnamefont
  {Nicolis}}, \bibinfo {author} {\bibfnamefont {R.}~\bibnamefont {Rattazzi}}, \
  and\ \bibinfo {author} {\bibfnamefont {E.}~\bibnamefont {Trincherini}},\
  }\href {\doibase 10.1007/JHEP05(2010)095} {\bibfield  {journal} {\bibinfo
  {journal} {JHEP}\ }\textbf {\bibinfo {volume} {05}},\ \bibinfo {pages} {095}
  (\bibinfo {year} {2010})},\ \bibinfo {note} {[Erratum: JHEP 11, 128
  (2011)]},\ \Eprint {http://arxiv.org/abs/0912.4258} {arXiv:0912.4258
  [hep-th]} \BibitemShut {NoStop}%
\bibitem [{\citenamefont {de~Rham}\ \emph {et~al.}(2017)\citenamefont
  {de~Rham}, \citenamefont {Melville}, \citenamefont {Tolley},\ and\
  \citenamefont {Zhou}}]{deRham:2017avq}%
  \BibitemOpen
  \bibfield  {author} {\bibinfo {author} {\bibfnamefont {C.}~\bibnamefont
  {de~Rham}}, \bibinfo {author} {\bibfnamefont {S.}~\bibnamefont {Melville}},
  \bibinfo {author} {\bibfnamefont {A.~J.}\ \bibnamefont {Tolley}}, \ and\
  \bibinfo {author} {\bibfnamefont {S.-Y.}\ \bibnamefont {Zhou}},\ }\href
  {\doibase 10.1103/PhysRevD.96.081702} {\bibfield  {journal} {\bibinfo
  {journal} {Phys. Rev. D}\ }\textbf {\bibinfo {volume} {96}},\ \bibinfo
  {pages} {081702} (\bibinfo {year} {2017})},\ \Eprint
  {http://arxiv.org/abs/1702.06134} {arXiv:1702.06134 [hep-th]} \BibitemShut
  {NoStop}%
\bibitem [{\citenamefont {Klauder}(1960)}]{Klauder:1960kt}%
  \BibitemOpen
  \bibfield  {author} {\bibinfo {author} {\bibfnamefont {J.~R.}\ \bibnamefont
  {Klauder}},\ }\href {\doibase 10.1016/0003-4916(60)90131-7} {\bibfield
  {journal} {\bibinfo  {journal} {Annals Phys.}\ }\textbf {\bibinfo {volume}
  {11}},\ \bibinfo {pages} {123} (\bibinfo {year} {1960})}\BibitemShut
  {NoStop}%
\bibitem [{\citenamefont {Glauber}(1963)}]{Glauber:1963tx}%
  \BibitemOpen
  \bibfield  {author} {\bibinfo {author} {\bibfnamefont {R.~J.}\ \bibnamefont
  {Glauber}},\ }\href {\doibase 10.1103/PhysRev.131.2766} {\bibfield  {journal}
  {\bibinfo  {journal} {Phys. Rev.}\ }\textbf {\bibinfo {volume} {131}},\
  \bibinfo {pages} {2766} (\bibinfo {year} {1963})}\BibitemShut {NoStop}%
\bibitem [{\citenamefont {Gubser}(2011)}]{Gubser:2011mp}%
  \BibitemOpen
  \bibfield  {author} {\bibinfo {author} {\bibfnamefont {S.~S.}\ \bibnamefont
  {Gubser}},\ }\href {\doibase 10.1016/j.physletb.2011.10.028} {\bibfield
  {journal} {\bibinfo  {journal} {Phys. Lett. B}\ }\textbf {\bibinfo {volume}
  {705}},\ \bibinfo {pages} {279} (\bibinfo {year} {2011})},\ \Eprint
  {http://arxiv.org/abs/1109.5687} {arXiv:1109.5687 [hep-th]} \BibitemShut
  {NoStop}%
\bibitem [{\citenamefont {Ramos}\ \emph {et~al.}(2020)\citenamefont {Ramos},
  \citenamefont {Spierings}, \citenamefont {Racicot},\ and\ \citenamefont
  {Steinberg}}]{Ramos:2019one}%
  \BibitemOpen
  \bibfield  {author} {\bibinfo {author} {\bibfnamefont {R.}~\bibnamefont
  {Ramos}}, \bibinfo {author} {\bibfnamefont {D.}~\bibnamefont {Spierings}},
  \bibinfo {author} {\bibfnamefont {I.}~\bibnamefont {Racicot}}, \ and\
  \bibinfo {author} {\bibfnamefont {A.~M.}\ \bibnamefont {Steinberg}},\ }\href
  {\doibase 10.1038/s41586-020-2490-7} {\bibfield  {journal} {\bibinfo
  {journal} {Nature}\ }\textbf {\bibinfo {volume} {583}},\ \bibinfo {pages}
  {529} (\bibinfo {year} {2020})},\ \Eprint {http://arxiv.org/abs/1907.13523}
  {arXiv:1907.13523 [physics.atom-ph]} \BibitemShut {NoStop}%
\end{thebibliography}%

\end{document}